# Vortex glass line and vortex liquid resistivity in doped BaFe$_2$As$_2$ single crystals

S. R. Ghorbani,[1,2] X.L. Wang,[1]* M. Shabazi,[1] S. X. Dou,[1] K.Y. Choi[3], and C.T. Lin[4]

[1]Institute for Superconducting and Electronic Materials, University of Wollongong, Wollongong, New South Wales 2522, Australia
[2]Department of Physics, Sabzevar Tarbiat Moallem University, P.O. Box 397, Sabzevar, Iran
[3]Frontier Physics Research Division and Department of Physics and Astronomy, Seoul National University, Seoul 151-747, Korea
[4]Max Planck Institute for Solid State Research, Heisenbergstr 1, 70569 Stuttgart, Germany

(Sep. 14, 2011)

The vortex liquid-to-glass transition has been studied in Ba$_{0.72}$K$_{0.28}$Fe$_2$As$_2$, Ba$_{0.9}$Co$_{0.1}$Fe$_2$As$_2$, and Ba(Fe$_{0.45}$Ni$_{0.05}$)$_2$As$_2$ single crystal with superconducting transition temperature, T$_c$ = 31.7, 17.3, and 18 K, respectively, by magnetoresistance measurements. For temperatures below T$_c$, the resistivity curves were measured in magnetic fields within the range of 0 ≤ B ≤ 13 T, and the pinning potential was scaled according to a modified model for vortex liquid resistivity. Good scaling of the resistivity ρ(B, T) and the effective pinning energy U$_0$(B,T) was obtained with the critical exponents s and B$_0$. The vortex state is three-dimensional at temperatures lower than a characteristic temperature T$^*$. The vortex phase diagram was determined based on the evolution of the vortex-glass transition temperature T$_g$ with magnetic field and the upper critical field, H$_{c2}$. We found that non-magnetic K doping results in a high glass line close to the H$_{c2}$, while magnetic Ni and Co doping cause a low glass line which is far away from the H$_{c2}$. Our results suggest that non-magnetic induced disorder is more favourable for enhancement of pinning strength compared to magnetic induced disorder. Our results show that the pinning potential is responsible for the difference in the glass states.

In the mixed states of type II superconductors, vortices form the Abrisokov lattice as a result of quantization of magnetic flux. Depending on the magnetic field, the vortex lattice can change from the solid to the glass state, and will further transform to a liquid state and disappear at H > H$_{c2}$, where H$_{c2}$ is the upper critical field. The boundary between the lattice and glass or the glass and liquid phases is strongly affected by both the anisotropy and the disorder of a superconductor. The vortex glass line can be significantly suppressed in highly anisotropic superconductors such as high superconducting transition temperature (T$_c$) cuprates, as the field required for the glass state is inversely proportional to the square of the anisotropy.[1] Local disorder is another important factor affecting the vortex glass line. In high-T$_c$ cuprates, vortex lattices are soft and form two-dimensional pancake vortices, which are easily melted into vortex liquid by magnetic field and thermal fluctuations. Therefore, the glass line is far below H$_{c2}$ in cuprates. From the viewpoint of applications, the glass line determines the critical field below which the vortices can be pinned or supercurrent can still survive. The Fe-based superconductors are found to show a small anisotropy in REFeAsO$_{1-x}$F$_x$ (RE-1111 phase, with RE a rare earth element)[2-9] and doped BaFe$_2$As$_2$ (122 phase).[10,11] It has been found that doping the latter with K, a non-magnetic dopant occupying Ba sites, results in a maximum T$_c$ of 35 K and induces a very strong intrinsic pinning strength with high critical current density (J$_c$) and H$_{c2}$ simultaneously.[10] Whereas, magnetic dopants, such as Co and Ni, lead to a low T$_c$ of about 18-22 K, and lower J$_c$ and H$_{c2}$ than for K doped 122, although the grain boundaries are not detrimental to the J$_c$, as has been reported for Co-doped 122 epitaxial thin films.[12] So far, little work has been reported on the determination of glass lines in the pnictide compounds. These facts have motivated us to raise a fundamental question as to whether or not disorder induced by magnetic or non-magnetic dopants can cause great differences in the glass lines, since the anisotropy is only around 2-3 in 122 compounds. In this work, the vortex liquid-to-glass transition have been studied in Ba$_{0.72}$K$_{0.28}$Fe$_2$As$_2$, Ba$_{0.9}$Co$_{0.1}$Fe$_2$As$_2$, and Ba(Fe$_{0.45}$Ni$_{0.05}$)$_2$As$_2$ single crystal with T$_c$ = 31.7, 17.3, and 18 K, respectively, by magnetoresistance measurements. We found that non-magnetic K doping results in a high glass line close to the H$_{c2}$, while magnetic Ni and Co doping cause a low glass line which is far away from the H$_{c2}$. Our results suggest that non-magnetic induced disorder is more favourable for enhancement of pinning strength compared to magnetic induced disorder.

The phase transition can be clearly visualized through the difference in resistive behaviour between the solid and liquid phases. The broadening of the resistivity transition in magnetic field is a direct consequence of the thermal fluctuation in the vortex system. Therefore, resistive transport measurements are commonly used to study vortices and vortex phase transitions.[1,13,17-19] The vortex phase transition in cuprate superconductors can be understood using vortex-glass theory. According to this theory,[13] in the vortex glass state and close to the glass transition temperature T$_g$, the resistivity decreases as a power law

$$\rho = \rho_0 \left| \frac{T}{T_g} - 1 \right|^s \quad (1)$$

where s is a constant related to the various types of disorder, which can be introduced by columnar defects, point defects, dopants, and vacancies. ρ$_0$ is the characteristic resistivity of the normal state. Eq. (1) can be modified by using the energy difference k$_B$T-U$_0$, where U$_0$ is the effective pinning energy. Therefore, the driving force of the transition depends on the energy difference, instead of the temperature difference of T-T$_g$.[15] Therefore, Eq. (1) is re-formulated as

$$\rho = \rho_n \left| \frac{k_B T}{U_0(B,T)} - 1 \right|^s \quad (2)$$

Here ρ$_n$ is the normal state resistivity at the onset of the transition. In this model, the vortex solid to vortex liquid transition occurs when the two energy scales are equal, U$_0$(B,T$_g$) = k$_B$T$_g$. An empirical effective pinning energy was found[18]

$$U_0(B,T) = U_B \left(1 - \frac{T}{T_c}\right) \quad \text{with} \quad U_B = k_B T_c / (B/B_0)^\beta \quad (3)$$

where both B$_0$, which is inversely proportional to the square of the mass anisotropy, and β are temperature and field-independent constants. By considering the effective pinning energy at the glass temperature, i.e. U$_0$(B,T$_g$) = k$_B$T$_g$, the temperature dependent vortex glass line is obtained from[14,15]

$$B_g(T) = B_0 \left(\frac{1-T/T_c}{T/T_c}\right)^{1/\beta} \quad (4)$$

This has been used to determined the vortex glass for Y-123.[14-17] A useful scaling form for resistivity was obtained by combination of Eqs. (2) and (3)

$$\rho = \rho_n \left|\frac{T(T_c - T_g)}{T_g(T_c - T)} - 1\right|^s \quad (5)$$

Therefore, the resistivity in the vortex glass state depends on magnetic field through $T_g$. Experimental evidence for the vortex-glass phase has been obtained for various high-$T_c$ superconductors such as Y-123,[14-16] Tl-2212,[17,18] $MgB_2$,[19] and $Ba_{0.55}K_{0.45}Fe_2As_2$ [20] through the scaling exponent, as predicted by vortex–glass and modified vortex–glass theory. However the vortex-glass phase transition and scaling behaviour have not been investigated systematically in detail in the Fe-based superconductors.

Here, we have studied the glass transition from the vortex liquid side in $Ba_{0.72}K_{0.28}Fe_2As_2$, $Ba_{0.9}Co_{0.1}Fe_2As_2$, and $Ba(Fe_{0.45}Ni_{0.05})_2As_2$ single crystal by magnetoresistance measurements. Good scaling of the $\rho(B, T)$ and $U_0(B,T)$ were obtained with the critical exponents $s$ for all crystals. The vortex phase diagram was determined based on the evolution of the vortex-glass transition temperature $T_g$ with magnetic field and the upper critical field. We have also compared the modified vortex-glass model and the thermally activated flux flow model in the vortex liquid region.

Single crystals with the nominal composition $Ba_{0.72}K_{0.28}Fe_2As_2$, $Ba_{0.9}Co_{0.1}Fe_2As_2$, and $Ba(Fe_{0.45}Ni_{0.05})_2As_2$ were prepared by a self-flux method. Details of the single crystal growth are described elsewhere.[21] The as-grown single crystals were cleaved and shaped into rectangular bars for measurements. The transport properties were measured over a wide range of temperature and magnetic fields up to 13 T using a physical properties measurement system (PPMS, Quantum Design).

The electrical resistance of $Ba_{0.72}K_{0.28}Fe_2As_2$ (BaK-122), $Ba_{0.9}Co_{0.1}Fe_2As_2$ (BaCo-122), and $Ba(Fe_{0.45}Ni_{0.05})_2As_2$ (BaNi-122) single crystals, measured at fields up to 13 T for H//c, is shown in the Arrhenius plot in Fig. 1. The transition temperatures, $T_c$, are 31.7, 17.3, and 18.0 K, with almost the same small transition width ($\Delta T_c$), of 0.7 ± 0.1 K for BaK-122, BaCo-122, and BaNi-122 single crystals, respectively.

According to the vortex glass model, Eq. (1), the resistance goes to zero at the glass temperature $T_g$ as $\rho \propto (T-T_g)^s$. Consequently $T_g(B)$ can be extracted by applying the Vogel-Fulcher relation, ($d \ln\rho/dT)^{-1} \propto (T-T_g)$, to the resistive tails. According to Eq. (5), the inverse of the logarithmic derivative is

$$\left(\frac{\partial ln\rho}{\partial T}\right)^{-1} = \frac{T-T_g}{s}\left(\frac{T_c-T}{T_c-T_g}\right). \quad (6)$$

This expression only differs from the Vogel-Fulcher relation by a correction factor $A(T) = (T_c-T)/(T_c-T_g)$ which is close to one at temperatures sufficiently close to $T_g$. Therefore, the usual Vogel-Fulcher relation can be applied to estimate $T_g$ directly. As shown in Fig. 2, the low temperature data are consistent with the Vogel-Fulcher relation up to the temperature of $T^*$, which shows a deviation from the straight line, with an intercept of $T_g$ for BaK-122, BaCo-122, and BaNi-122. At temperature $T^*$, vortices change from 2D to 3D. In layered superconductors above this temperature, the vortex flux lines act as a two-dimensional, 2D, pancake vortex.[22]

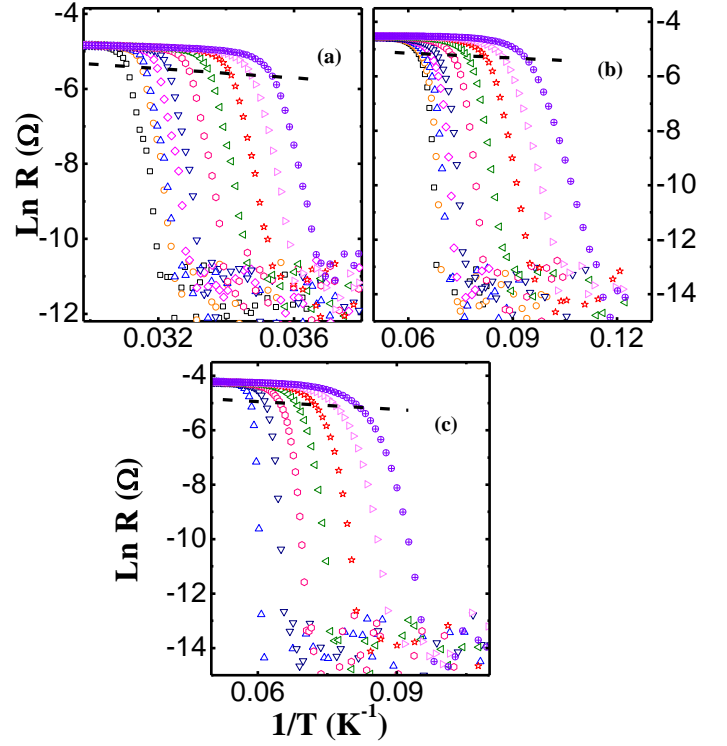

Fig. 1. Arrhenius plot of the resistance for (a) $Ba_{0.72}K_{0.28}Fe_2As_2$ (BaK-122), (b) $Ba_{0.9}Co_{0.1}Fe_2As_2$ (BaCo-122), and (c) $Ba(Fe_{0.45}Ni_{0.05})_2As_2$ (BaNi-122) single crystals for (from left to right) B = 0.1, 0.5, 1, 2, 3, 5, 7, 9, 11, and13 T. The upper limit for the applicability of the Vogel-Fulcher relation is marked by the dashed line.

The obtained $s$, the inverse of the slope of the straight line, is 6.0 ± 0.5, 5.3 ± 0.3 and 5.1 ± 0.4 for BaK-122, BaCo-122, and BaNi-122 single crystals, respectively. For BaK-122, the $s$ parameter is roughly three times larger than that obtained for $Ba_{0.55}K_{0.45}Fe_2As_2$.[20]

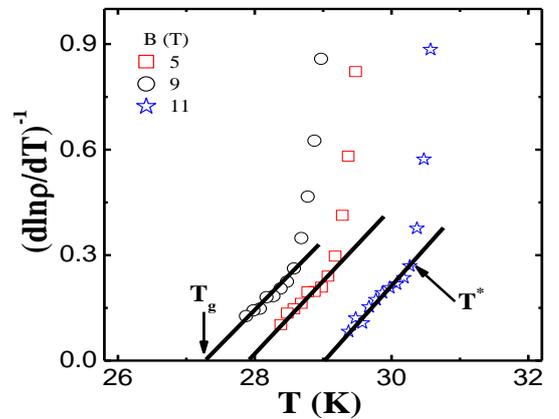

Figure 2. Determination of the glass transition temperature from the Vogel-Fulcher relation for BaK-122 at three different magnetic fields.

Equation (5) suggests a scaling behaviour of the vortex glass between the normalized resistivity $\rho/\rho_n$ and the scaled temperature $T_{sc} = [T(T_c-T_g)/T_g(T_c-T)]-1$. Figure 3 shows the $\rho/\rho_n$ versus scaling temperature $T_{sc}$. As can be seen in Fig. 3, the resistivity transition at fields between 0 and 13 T has been scaled into one curve with the obtained critical exponent Vortex glass with the same resistivity scaling behaviour has been observed in several high-$T_c$ superconductors such as oxygen- deficient Y-123 single crystal[15,16] and Tl-2212 thin film.[17]

The upper critical field, $B_{c2}$, is obtained from the 90% values of its corresponding resistivity transition. Using the estimated vortex glass line $B_g$, the B-T phase diagram for the BaK-122, BaCo-122, and BaNi-122 single crystals is shown in Fig. 4. According to the collective pinning model,[23] the disorder-induced spatial fluctuations in the solid-vortex lattice can be clearly divided into markedly different regimes according to the strength of the applied field. As is shown in the inset of Fig. 4, two different regimes are distinguishable: (1) vortex glass, which governs the region below $B_g$; (2) vortex liquid, which holds between $B_g$ and $B_{c2}$, where thermal fluctuations are important. As can be seen from Fig. 4, the vortex-glass phase indicates that the K doped BaFe$_2$As$_2$ single crystal has a very narrow region of the vortex-liquid phase, which is denoted by ΔT in the inset of

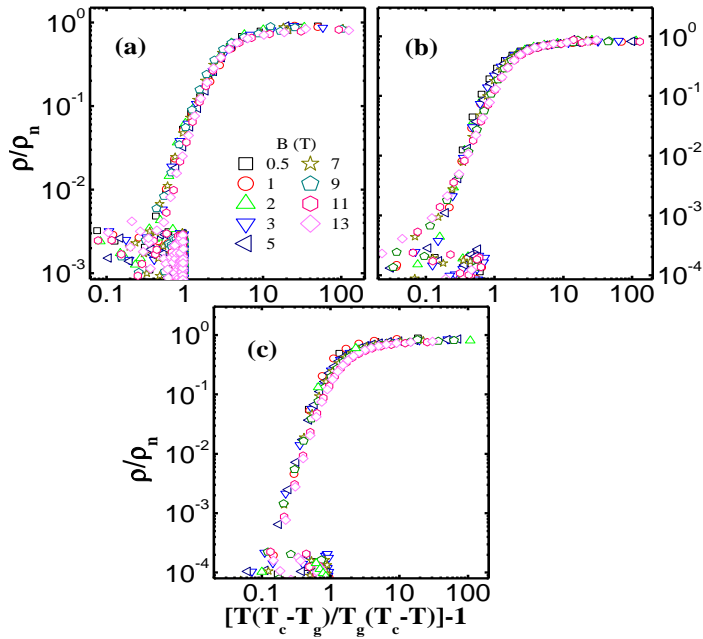

Fig. 3. Resistivity scaling according to Eq. (5) for (a) Ba$_{0.72}$K$_{0.28}$Fe$_2$As$_2$, (b) Ba$_{0.9}$Co$_{0.1}$Fe$_2$As$_2$, and (c) Ba(Fe$_{0.45}$Ni$_{0.05}$)$_2$As$_2$ single crystal for $0 \leq B \leq 13$ T.

Figure 4, with a small ΔT of 0.06 ± 0.02 T and at magnetic field of 0.5 up to 13 T (weakly field dependent), which has originated from the vastly enhanced vortex pinning. However, the glass lines for both Co (ΔT = 0.18 ± 0.02) and Ni doped 122 single crystals are far from the $H_{c2}$ lines and strongly temperature dependent for Ni122 (ΔT = 0.03–0.22 for the B = 0.5–13 T range). This indicates that the vortex liquid state exists in a wide range of fields below $H_{c2}$. This is significantly different from BaK-122. The result shows that the vortex liquid region for BaK-122 and BaCo-122 is roughly magnetic field independent (ΔT), while it does depend on temperature for BaNi-122 because ΔT increases with increasing magnetic field.

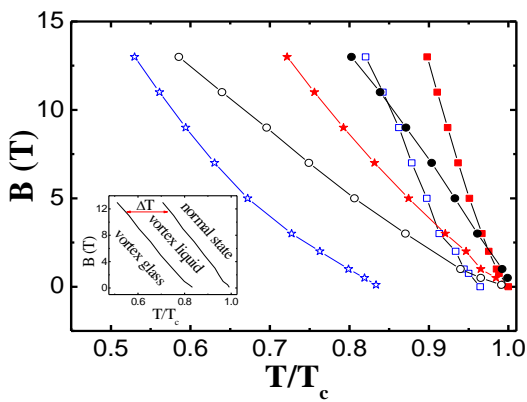

Fig. 4. Phase diagrams of Ba$_{0.72}$K$_{0.28}$Fe$_2$As$_2$ (squares), Ba$_{0.9}$Co$_{0.1}$Fe$_2$As$_2$ (stars), and Ba(Fe$_{0.45}$Ni$_{0.05}$)$_2$As$_2$ (circles) single crystals. $B_g$ was obtained from the experimental ρ(T) and $U_0$(T) data (see Figs. 2 and 4). $B_{c2}$(T) values were obtained from the criterion of the 90% values of their corresponding resistivity transition. Open symbols: $B_{c2}$ and solid symbols: $B_g$. Inset: general phase diagram.

Another important point is that one can directly observed that ΔT of BaK-122 is smaller than for BaCo-122 or BaNi-122, which is an indication that the pinning potential of BaK-122 is stronger than for either BaCo-122 or BaNi-122 single crystals. This is supported by the pinning potential results, as can be seen in Fig. 6. The results show that non- magnetic doping, such as K on Ba site, and magnetic doping, such as Co or Ni on the Fe site, have different effects on the vortex glass line. The question is what the effect will be on the vortex glass line of magnetic and non-magnetic substitution on the same site, especially the Fe site, which is located on the FeAs plane. This requires further study.

By solving Eq. (2) for $U_0$, one obtains

$$U_0(B,T) = k_B T \left[1 + \left(\frac{\rho}{\rho_n}\right)^{1/s}\right]^{-1} \qquad (7)$$

According to Eq. (7), the pinning potential can be calculated directly from experimental data, provided that one knows the $\rho_n$ and the exponent s. The s parameter is estimated as the inverse slope of the resistivity in the vortex glass state, as mentioned above, and the $\rho_n$ has been taken as the normal state resistivity at T = 35 K for BaK-122 and T= 20 K for both BaCo-122 and BaNi-122 single crystals. The calculated $U_0$(B,T) of the BaK-122, BaCo-122, and BaNi-122 single crystals for fields between 0 and 13 T is shown in Fig. 5. As clearly shown by the solid lines, the low resistivity of the curves is well described by Eq. (3) with a field dependent $U_B$.

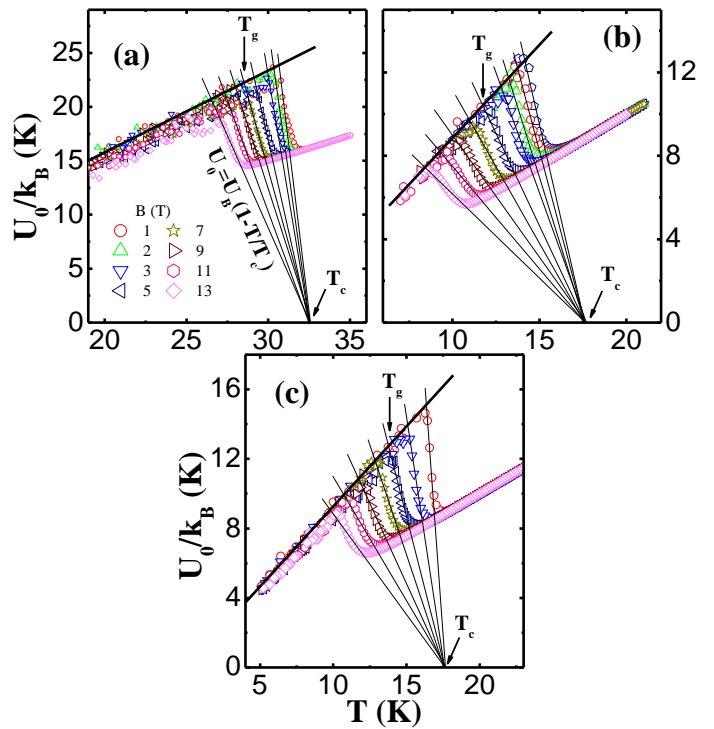

Fig. 5. The pinning potential $U_0$(B,T) as calculated according to Eq. (7) in field of $0 \leq H \leq 13$ T for (a) Ba$_{0.72}$K$_{0.28}$Fe$_2$As$_2$, (b) Ba$_{0.9}$Co$_{0.1}$Fe$_2$As$_2$, and (c) Ba(Fe$_{0.45}$Ni$_{0.05}$)$_2$As$_2$ single crystal.

As can be seen in Fig. 5, an extrapolation of the linear behaviour of $U_0$(T) at temperatures close to $T_g$ and at different fields merges in the point T = $T_c$ and $U_0$ = 0 which is in good agreement with Eq. (3). Therefore, the field dependence of the pinning energy $U_B$ can be found directly from the slope of these lines. The magnetic dependence of the estimated result from Fig.

5 for $U_B$ is shown in Fig. 6. As can be seen in Figure 6, the pinning potential of $Ba_{0.72}K_{0.28}Fe_2As_2$ single crystal is constant, while a different power law dependence is found at low and high magnetic fields for both $Ba_{0.9}Co_{0.1}Fe_2As_2$, and $Ba(Fe_{0.45}Ni_{0.05})_2As_2$ single crystal. The $U_B/k_B$ decreases slowly with increasing applied magnetic field for B < 3 T, scaled as $B^{-0.06}$, and then decreases as $B^{-0.9}$ for B > 3 T. This result suggests that the single vortex pinning may co-exist with collective creep in low fields and then the collective creep dominates in high magnetic fields.

Another important point is that one can directly obtain $T_g$ by considering the crossing points of the lines $U_0(B,T)$ and the line $U_0 = k_BT$, as indicated by the arrows in Fig. 5. This criterion was used for estimating $T_g$ in the modified vortex-glass model.[15,16] Therefore, $U_0(B,T)$ is the average pinning energy in the system and is responsible for the vortex solid to liquid transition when $U_0(B,T_g) = k_BT_g$, as discussed in the Introduction.

In many studies, the vortex liquid resistivity is described by a thermally activated flux flow (TAFF) model, $\rho(T, H) = \rho_n \exp(U^*/k_BT)$,[24] where $\rho_n$ is the normal state resistivity and $k_B$ is the Boltzmann constant. By using the TAFF model, we obtained the activation energy, $U^*/k_B = \partial \ln\rho/\partial(1/T)$, which is shown in the inset of Fig. 6. At B = 1 T, $U^*/k_B$ was 6400, 3311, and 1626 K for the BaK-122, BaCo-122, and BaNi-122 single crystals, respectively, while $U_B/k_B$ was 192, 77, and 32 K. Although $U^*$ is larger than $U_B$ in value, it has a similar magnetic field dependence.

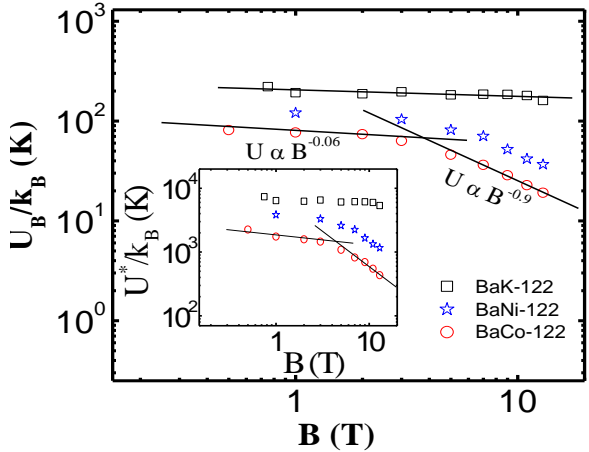

Fig. 6. Magnetic field dependence of $U_B/k_B$ as estimated from Eq. (3) and the slopes in Fig. 4. The solid lines are a fit to the data giving the relation $U_B/k_B \approx B^{-n}$ with n = 0.06 at B < 3 T and n = 0.9 at B > 3 T. Inset: Field dependence of the estimated activation energy based on the thermally activated flux flow (TAFF) model.

For comparison of these values, by taking the derivative $\partial \ln\rho/\partial(1/T)$ of Eq. (2) and using Eq. (3), one can obtain the relation

$$U^* = \frac{s[1+(\rho/\rho_n)^{1/s}]^2}{(\rho/\rho_n)^{1/s}} U_B = AU_B \qquad (8)$$

By using *s* and experimental data of $\rho/\rho_n \approx 10^{-1}$-$10^{-3}$, the correction factor A is between 24-33, 21-32, and 20-28 for the BaK-122, BaCo-122, and BaNi-122 single crystals, respectively. Therefore, there is an excellent agreement between both models in the vortex liquid region.

In conclusion, it was shown that the glass transition introduced based on a modified model for the vortex-glass transition can be applied to the doped $BaFe_2As_2$ superconductors. For temperature below the superconducting transition temperature, a scaling of all measured resistivity $\rho(B, T)$ and pinning potential $U_0(B,T)$ values in magnetic fields up to 13 T with the critical exponents *s* is obtained. The vortex phase diagram was determined based on the evolution of the vortex-glass transition temperature $T_g$ with magnetic field and the upper critical field. The results suggest that the vortex region of $Ba_{0.72}K_{0.28}Fe_2As_2$ is shorter than it is for $Ba_{0.9}Co_{0.1}Fe_2As_2$ and $Ba(Fe_{0.45}Ni_{0.05})_2As_2$, and it is also magnetic field independent in magnetic fields smaller than 13 T. It was found that the pinning force of $Ba_{0.72}K_{0.28}Fe_2As_2$ is stronger than for $Ba_{0.9}Co_{0.1}Fe_2As_2$ and $Ba(Fe_{0.45}Ni_{0.05})_2As_2$. Furthermore, we compared these results with the thermally activated flux flow behaviour which is usually employed to account for the resistivity in the vortex liquid region.

## ACKNOWLEDGMENTS

This work was supported by the Australian Research Council through an ARC discovery project DP1094073.

*Corresponding author
E-mail: xiaolin@uow.edu.au